# Many-to-One Throughput Capacity of IEEE 802.11 Multi-hop Wireless Networks

Chi Pan Chan, *Student Member, IEEE,* Soung Chang Liew, *Senior Member IEEE*, and An Chan, *Student Member, IEEE*

**Abstract**—This paper investigates the many-to-one throughput capacity (and by symmetry, one-to-many throughput capacity) of IEEE 802.11 multi-hop networks.  It has generally been assumed in prior studies that the many-to-one throughput capacity is upper-bounded by the link capacity $L$. Throughput capacity $L$ is not achievable under 802.11. This paper introduces the notion of "canonical networks", which is a class of regularly-structured networks whose capacities can be analyzed more easily than unstructured networks. We show that the throughput capacity of canonical networks under 802.11 has an analytical upper bound of $3L/4$ when the source nodes are two or more hops away from the sink; and simulated throughputs of $0.690L$ ($0.740L$) when the source nodes are many hops away. We conjecture that $3L/4$ is also the upper bound for general networks. When all links have equal length, $2L/3$ can be shown to be the upper bound for general networks. Our simulations show that 802.11 networks with random topologies operated with AODV routing can only achieve throughputs far below the upper bounds. Fortunately, by properly selecting routes near the gateway (or by properly positioning the relay nodes leading to the gateway) to fashion after the structure of canonical networks, the throughput can be improved significantly by more than 150%. Indeed, in a dense network, it is worthwhile to deactivate some of the relay nodes near the sink judiciously.

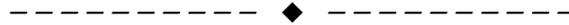

## 1 INTRODUCTION

Many-to-one communication is a common communication mode in many multi-hop wireless networks. Two relevant applications are sensor networks and multi-hop wireless mesh networks. In sensor networks, there is often a "data processing center" to which data collected at distributed sensors are to be forwarded. In multi-hop wireless mesh networks, there is an Internet gateway connecting the mesh network to the core wired Internet – the client stations and the Internet gateway form a many-to-one relationship.

This paper investigates the many-to-one throughput capacity of IEEE 802.11 multi-hop networks. In this setting, there are multiple source nodes generating traffic streams to be forwarded to a common sink node via relay nodes. The relay nodes could be sources themselves. By symmetry, the throughput capacity thus found is also the same as that in a one-to-many scenario in which a source node generates multiple *distinct* data streams to be forwarded to their respective sinks (note: this is not to be confused with the multicast scenario in which the same data is to be forwarded to multiple sinks). For convenience, we shall refer to the sink in the many-to-one scenario as the "center" of the network.

There have been many related studies on the capacity of general wireless networks. Gupta and Kumar [1] analyzed the capacity in many-to-many situation. It provides the basic model that can be adapted for use in the analysis of the many-to-one communication. As a loose bound, it is obvious that the many-to-one throughput capacity is upper-bounded by $L$ [1]-[3], where $L$ is the single-link throughput capacity, since this is the rate at which the sink can receive data. There is a high probability, however, that the throughput capacity is lower than $L$ for a random network [3]. This paper follows the approach used in [1]-[3] in characterizing which nodes can transmit together without packet collisions. The main difference is that here we are interested in the capacity throughput obtained under the IEEE 802.11 distributed MAC protocol [4]. Specifically, we integrate into our analysis the effects of carrier sensing,  the existence of an ACK frame for each DATA frame transmission, and the distributed nature of the CSMA protocol,  while [1]-[3] do not and their bounds are obtained with the implicit assumption of perfectly scheduled transmissions.

There are three main contributions to this paper:

1. We introduce the notion of "canonical networks", which is a class of regularly-structured networks whose capacities can be analyzed more easily than general unstructured networks. We find that the throughput capacity of canonical networks under 802.11 is upper bounded by $3L/4$ when the source nodes are at least two hops away from the sink. We conjecture that this is also the upper bound for general networks. Indeed, when all the links in the network are of equal length, canonical networks and general networks have the same upper bound of $2L/3$.

————————————————
- *All authors are with the Department of Information Engineering, The Chinese University of Hong Kong, New Territories, Hong Kong. E-mail: C.P. Chan : cpchan4@ie.cuhk.edu.hk , S. C. Liew : soung@ie.cuhk.edu.hk, A. Chan : achan5@ie.cuhk.edu.hk.*

*Manuscript received (insert date of submission if desired). Please note that all acknowledgments should be placed at the end of the paper, before the bibliography.*

2. We find that canonical networks give much insight on how a many-to-one network should be designed in general. Our simulations show that 802.11 networks with random topologies operated with AODV routing can only achieve throughputs far below the upper bound of canonical networks. However, if we route the traffic in accordance to the optimized routes obtained from an optimization algorithm, the routes near the center have a structure similar to that of the optimal canonical network structure. In other words, as a principle, routing or network design near the center should be fashioned after the canonical network. Our further investigation shows that a "*manifold*" *canonical network* structure near the center may yield throughput improvement of more than 150% relative to that obtained by using AODV routing in a general network structure. Indeed, in a dense network, it is worthwhile to deactivate some of the relay nodes near the sink judiciously.

3. We find that ensuring the many-to-one network is hidden-node free (HNF) in our design leads to higher throughputs as compared to not doing so. This is in contrast to the many-to-many case, in which the large carrier-sensing range required to ensure the HNF property may lower the network throughput due to the increased exposed-node problem [5]. This observation is used as a design principle in much of the study in 1 and 2 above.

The rest of this paper is organized as follows. Section II provides the definitions and assumptions used in our analysis. Section III derives the throughput capacities of canonical networks, and presents simulation results to support our findings. In addition, we demonstrate the desirability of ensuring the HNF property in many-to-one networks. Section IV investigates general networks not restricted to the canonical network structure. We show that the optimal routing in general networks results in a subset of selected routes that form a structure near the center that resembles the optimal canonical network. We then apply this insight to demonstrate the desirability of designing the network according to a "manifold" canonical-network structure near the center. Section V concludes this paper.

## 2 DEFINITIONS AND ASSUMPTIONS

Let us first provide some definitions used in our analysis.

*Definition 1*: The *source nodes* are nodes that generate data traffic.

*Definition 2*: The *sink node* is the center to which the data collected at the source nodes are to be forwarded.

*Definition 3*: The *relay nodes* relay data traffic from the source nodes to the sink node.

Note that a node can be classified as one of the followings: 1) a source node; 2) a sink node; 3) a relay node; or 4) both a source node and a relay node.

*Definition 4*: Given a network topology, the *uniform throughput capacity* $C^u$ with respect to a set of source nodes and a sink node is the maximum total rate at which the data can be forwarded to the sink node, with equal amount of traffic from each source node to the sink node. The *throughput capacity*, $C^m$, on the other hand, does not require equal amounts of traffic from sources to sink. Thus, in general, $C^m \geq C^u$.

Fig. 1 shows a simple example of a network consisting of three nodes. Suppose that node 2 is the sole source node and node 1 is the relay node that forwards packets from node 2 to node 0. Node 1 does not generate traffic by itself. Then, $C^m = L/2$, where $L$ is the capacity of one link. This is because node 1 cannot receive and transmit at the same time (typical assumption of half-duplexity of wireless links). Also, since there is only one source node, $C^m = C^u$.

If node 1 is also a source node in addition to being a relay node, then $C^m = L$ (obtained when only node 1 is allowed to transmit), and $C^u = 2L/3$, with nodes 1 and 2 having a throughput of $L/3$ each. Since node 1 needs to serve as the relay node for node 2, node 1 will need to transmit twice as often as node 2. So, proper scheduling is required.

Now, if we generalize the above linear network [7] to the one consisting ($n$+1) nodes, in which there are $n$ sources nodes with ($n$-1) of them also being relay nodes. Then, $C^u$ can be obtained as follows. Node 1 will transmit to node 0, the sink node, at rate $C^u$. Node 2 will transmit to node 1 at rate $C^u(n-1)/n$, and so on. In general, node ($i$+1) transmits to node $i$ at rate $C^u(n-i)/n$. We note that when node $i$ transmits, nodes ($i$+1) and ($i$+2) cannot: node ($i$+2) cannot transmit because the reception at node ($i$+1) will be corrupted by the transmission by node $i$. So, considering transmissions of nodes 3, 2, and 1 (which form the bottleneck), we have $\sum_{i=1}^{3} C^u(n-i+1)/n = L$. That is, $C^u = Ln/(3n-3) \approx L/3$ for large $n$.

We note that $L/3$ is also the $C^m$ if node $n$ were the only source node. As a matter of fact, $C^u = C^m = L/3$ if the source nodes in the linear network were nodes $i$ for $i \geq 3$ only. Thus, for reasonably large $n$, if the traffic from nodes 1 and 2 is only a small fraction of the total traffic to the sink, we could treat nodes 1 and 2 as pure relay, non-source, nodes. Once we do that, we then do not have to distinguish between $C^u$ and $C^m$.

We next consider a general many-to-one network, such as that in Fig. 2. *For the study of many-to-one networks in this paper, we focus on the case where the source nodes are two or more hops away from the sink.* This is a good approximation when the nodes within one hop to the sink only generate a small fraction of the total traffic.

*Definition 5*: The throughput capacity with respect a multi-access protocol $p$ (e.g., *IEEE 802.11*), $C_p$, is the total rate at which the data can be forwarded to the sink node using that protocol, assuming the source nodes are two or more hops away from the sink. The transmission

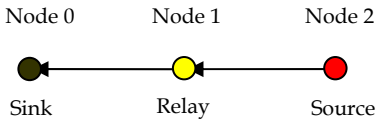

Fig. 1. Simple network example.

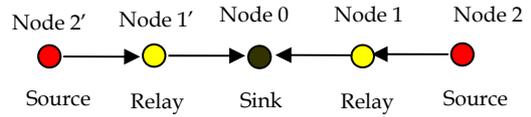

Fig. 3. A two-chain many-to-one network with equal link length.

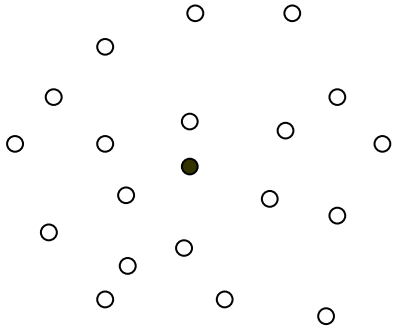

Fig. 2. A random many-to-one network.

schedule by the links is dictated by the protocol.

This paper focuses on the throughput capacity under the 802.11 CSMA protocol, $C_{802.11}$. Henceforth, by throughput capacity, we mean $C_{802.11}$. For illustration, let us consider the two-chain linear topology shown in Fig. 3. Suppose that only nodes 2 and 2' are the source nodes. Under "perfect scheduling", nodes 1 and 2' will transmit together; and nodes 1' and 2 will transmit together. This results in a throughput capacity of $L$. Under 802.11, however, the transmissions are usually not perfectly aligned in time. In addition, a DATA frame is followed by an ACK frame in the reverse direction. Suppose nodes 1 and 2' transmit together. Say, the transmission of the DATA frame of node 1 completes first, while the transmission node 2' is ongoing. When node 0 returns an ACK to node 1, this ACK also reaches node 1', the receiver of the transmission from node 2', causing a collision there. Thus, under 802.11, simultaneous transmissions by nodes 1 and 2' will usually result in a collision unless the completion times of their DATA transmissions are perfectly aligned, which is rare. In this case, $C_{802.11}$ is at best $2L/3$, since at best node 2 and 2' can transmit together, and nodes 1 and 1' will need to transmit at separate times.

For many-to-one networks, the capacity bottleneck is likely to be near the sink node because all traffic travels toward the sink node. Specifically, nodes near the sink node are responsible for forwarding more traffic, and these nodes contend for access of the wireless medium because they are close to each other. To obtain an idea on the upper limit of the throughput capacity under 802.11, we consider a class of networks referred to as the canonical networks. An example of a canonical network is shown in Fig. 4. We show that $3L/4$ is the upper bound of the throughput capacity of canonical networks, and conjecture that this is also the upper bound for networks with general structures. We will motivate the study of the canonical networks shortly. In the special case in which all links have equal length, then the throughput capacities of the canonical network as well as general networks are upper-bounded by $2L/3$. We now define the canonical networks.

*Definition 6*: A *chain* is formed by a sequence of *at least three* nodes leading to the center sink node. Traffic is forwarded from one node to the next node in the sequence on its way to the sink node. A *linear* chain is a chain which is a straight line.

In Fig. 4, for example, there are eight linear chains.

*Definition 7*: An *i-hop node* is a node that is $i$ hops away from the sink node in a chain (see Fig. 4).

*Definition 8*: A *canonical network* is formed by a number of linear chains leading to a *common* center sink node; the nodes in different chains are distinct except the sink node. In addition, the distance between an *i*-hop node and an (*i*-1)-hop node, $d_i$, is the same for all the linear chains (see Fig. 4).

*Definition 9*: A *ring* is a circle centered on the sink node. An *i*-hop ring consists of all the *i*-hop nodes of the different linear chains in a canonical network (see Fig. 4).

*Motivation for the Study of Canonical Networks*

Canonical networks have regular structures and can be analyzed more easily than general networks. We conjecture that the upper bound of throughput capacity obtained for canonical networks is also the upper bound for general networks, because intuitively canonical networks model a rich class of networks the optimal of which may yield very good throughput performance. Consider the following intuitive argument. (i) In a densely populated network (say, infinitely dense), we may choose to form linear chains from the source nodes to the center sink node for routing purposes. Since the direction of traffic flow is pointed exactly to the center, there is no "wastage" with respect to the case in which the routing direction is at an angle to the center. (ii) We have defined the class of canonical networks to be quite general in that we do not restrict the number of linear chains in it. Neither do we limit the distance $d_i$. In deriving the capacity of the canonical network later, we allow for the possibility of an infinite number of linear chains and arbitrarily small $d_i$. This provides us with a high degree of freedom in identifying the best-structured canonical networks. The above intuitive reasoning will be validated by simulation results later. In addition, we will show later that in a random network with many nodes (so that there is a high degree of freedom in forming routes), establishing a canonical-network-like structure near the center for routing purposes will generally lead to superior throughput performance.

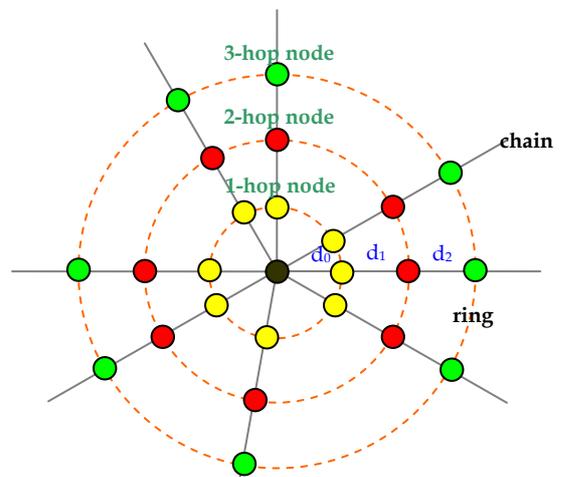

Fig. 4. A Canonical Network.

In this paper, unless otherwise stated, we further assume the following:

*Assumptions:*

(1) The nodes and links are homogenous. They are configured similarly, i.e., same transmission power, carrier-sensing range (CSRange), transmission rate, etc.

(2) ACK is sent by the receiver when a packet is received successfully, as per the 802.11 DCF operation.

(3) The following constraints apply to simultaneous transmissions [1][6]. Consider two links ($T_1$, $R_1$) and ($T_2$, $R_2$). For simultaneous transmissions without collisions, they must satisfy *all* the eight inequalities below:

$$|X_{T2} - X_{R1}| > (1+\Delta)|X_{T1} - X_{R1}|$$
$$|X_{R2} - X_{R1}| > (1+\Delta)|X_{T1} - X_{R1}|$$
$$|X_{T2} - X_{T1}| > (1+\Delta)|X_{T1} - X_{R1}|$$
$$|X_{R2} - X_{T1}| > (1+\Delta)|X_{T1} - X_{R1}| \quad (1)$$
$$|X_{T1} - X_{R2}| > (1+\Delta)|X_{T2} - X_{R2}|$$
$$|X_{R1} - X_{R2}| > (1+\Delta)|X_{T2} - X_{R2}|$$
$$|X_{T1} - X_{T2}| > (1+\Delta)|X_{T2} - X_{R2}|$$
$$|X_{R1} - X_{T2}| > (1+\Delta)|X_{T2} - X_{R2}|$$

where $X_i$ is the location of node $i$, $|X_i - X_j|$ is the distance between $X_i$ and $X_j$, $\Delta > 0$ is the distance margin (see next paragraph). These are the physical constraints that prevent DATA-DATA, DATA-ACK and ACK-ACK collisions.

The received power function can be expressed in the form of

$$P(d) \propto P_t / d^\alpha, \quad (2)$$

where $P_t$ is the transmission power, $d$ is the distance and α is the path-loss exponent, which typically ranges from 2 to 6 according to different environments [8]. By the assumptions that all the nodes have the same transmission power and α = 4, and Signal-to-Interference Ratio (SIR) requirement of 10dB. Then at $R_1$, we require

$$\frac{P(|X_{T1} - X_{R1}|)}{P(|X_{T2} - X_{R1}|)} > SIR \quad (3)$$

giving

$$\frac{|X_{T2} - X_{R1}|}{|X_{T1} - X_{R1}|} > \sqrt[4]{10} = 1.78$$

In other words, Δ = 0.78. Unless otherwise stated, we assume Δ = 0.78 throughout this work.

(4) In 802.11 networks, there are two types of packet collisions: collisions due to hidden nodes (HN) (see explanation of assumption (5) below or [6]), and collisions due to simultaneous countdown to zero in the backoff period of the MAC of different transmitters. In much of our throughput-capacity analysis, we will neglect the latter collisions and assume that they have only small effects toward throughput capacity, a fact which has been borne out by simulations and which can be understood through intuitive reasoning, particularly for a network in which a node is surrounded by only a few other *active nodes* who may collide with it. As will be shown later in this paper, this is generally a characteristic of a network with good throughput performance (see results of Fig. 14 and Fig. 18, for example). Also, an upper bound on throughput capacity obtained by neglecting the countdown collisions is still a valid upper bound. It is a good upper bound so long as it is tight. We will see later that the upper bounds we obtain are reasonably tight when verified against simulations results in which countdown collisions are taken into account. In the remainder of this paper, unless otherwise stated, the term "collisions" refers to collisions due to HN (i.e., caused by the failure of carrier-sensing) rather than simultaneous countdown to zero.

(5) In this paper, unless otherwise specified, we assume the so-called Hidden-Node Free Design (HFD) [6] in the network. That is, we design the network such that simultaneous transmissions that will cause collisions can be carrier-sensed by transmitters and be avoided. A reason for this assumption is that for many-to-one communication, eliminating hidden nodes is worthwhile (see simulation results in Section III-C). According to [6], HFD requires

(i) Use of Receiver Restart (RS) Mode, and
(ii) Sufficiently large CSRange.

This paper assumes the 802.11 basic mode and RTS/CTS are not used. We briefly describe the HFD requirements for understanding of the analysis later. More details can be found in [6]. Fig. 5 is an example showing that no matter how large CSRange is, the hidden node (HN) phenomenon can still occur in the absence of RS. In the figure, $T_1$ and $T_2$ are more than CSRange apart, and so simultaneous transmissions can occur. Furthermore, the SIR is sufficient at $R_1$ and $R_2$ so that no "physical collisions" occur. But HN can still happen, as described below.

Assume $T_1$ starts first to transmit a DATA packet to $R_1$. After the physical-layer preamble of the packet is received by $R_2$, $R_2$ will "capture" the packet and will not attempt to receive another new packet while $T_1$'s DATA is ongoing. If at this time $T_2$ starts to transmit a DATA to $R_2$, $R_2$ will not receive it and will not reply with an ACK to $T_2$, causing a transmission failure on link $(T_2, R_2)$. This is the default receiver mode assumed in the NS-2 simulator [10] and most 802.11 commercial products. Note that the example in Fig. 5 is independent of the size of CSRange.

This HN problem can be solved with the Receiver Restart Mode (RS) which can be enabled in some 802.11 products (e.g., Atheros Wi-Fi chips; however, the default is that this mode is not enabled). With RS, a receiver will switch to receive the stronger packet if its power is $C_t$ times greater than the current packet (say, 10 dB higher). The example in Fig. 5 will not give rise to HN with RS if CSRange is sufficiently large.

RS Mode alone, however, cannot prevent HN without sufficiently large CSRange. To see this, consider the example in Fig. 6. Assume $T_1$ transmits a DATA to $R_1$ first. During the DATA's period, $T_2$ starts to send a shorter DATA packet to $R_2$. With RS Mode, $R_2$ switches to receive $T_2$'s DATA and sends an ACK after the reception. If $T_1$'s DATA is still in progress, $R_2$'s ACK will corrupt the DATA at $R_1$, since the distance between $R_1$ and $R_2$ is within interference range $((1+\Delta)d_{max})$. To prevent $T_2$ from transmission (hence the collision), the following must be satisfied:

$$|X_{T1} - X_{T2}| \leq CSRange. \qquad (4)$$

Reference [6] proved that in general if $CSRange > (3+\Delta)d_{max}$, where $d_{max}$ is the maximum link length, then HN can be prevented in any network. However, for a specific network topology, e.g., the canonical network, the required CSRange can be smaller.

Throughout this work, we primarily focus on the pair-wise-interference model [1][6]. The concept of CSRange and the constraints in (1) rely on this assumption. An analysis which at the outset takes into account the simultaneous interferences from more than one source will complicate things significantly. So, given a network topology, our approach is to first identify the capacity based on pair-wise interference analysis only, and then verify the capacity is still largely valid under multiple interferences (this verification is done in Section III-D).

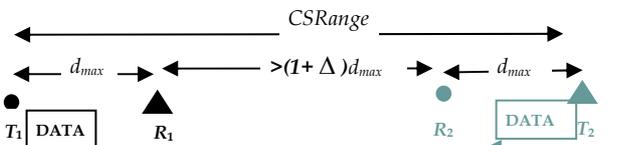

Fig. 5. Lack of RS Mode leads to HN no matter how large CSRange and SIR are.

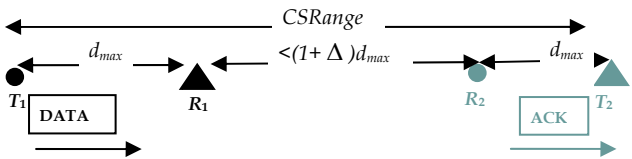

Fig. 6. With RS Mode, CSRange not sufficiently large still leads to HN due to insufficient SIR.

## 3 CANONICAL NETWORKS

In this section, we derive the throughput capacities of canonical networks. Section A analyzes two kinds of canonical networks: equal link-length and variable link-length networks. Simulation results are presented and discussed in Section B. Section C compares the performance of HFD and non-HFD networks, and Section D verifies the results under multiple interferences.

### 3.1 Theoretical Analysis

*(1) Equal Link-Length Networks*

We first consider the case where all links have the same length $d$, i.e., $d_0 = d_1 = \ldots = d$. Theorem 1, which follows from Lemma 1 and Corollary 1 below, proves that the throughput capacity in this network is upper-bounded by $2L/3$, where $L$ is the single-link throughput.

*Lemma 1*: Given three nodes on the periphery of a circle of radius $d$, we can identify two nodes with distance smaller than $(1+\Delta)d$ between them.

*Proof*: The three nodes form the vertices of a triangle. Consider the *equilateral* triangle inscribed on the circle of radius $d$, and let $t$ be the length of one side (see Fig. 7). Then

$$t = 2d \cos\frac{\pi}{6} = 1.731\, d < (1+\Delta)d$$

That is, it is not possible to inscribe a triangle with all sides no less than $(1+\Delta)d$ on the circle.

∎

*Corollary 1*: At any time, at most *two* 2-hop nodes can transmit at the same time.

*Proof*: With reference to Fig. 8, suppose that three 2-hop nodes can transmit together. In order that the ACK of any 1-hop node to not interfere with the reception of DATA packet of another transmission, the distances between the three 1-hop nodes must all be larger than $(1+\Delta)d$. By *Lemma 1*, this is not possible.

*Theorem 1*: For equal-link-length canonical networks, $C_{802.11} \leq 2L/3$, where $L$ is the link capacity.

*Proof*: Define "airtime" usage of a node to include the transmission time of DATA packets as well as the ACK from the receiver [7]. Let $S_{ij}$ be the airtime occupied by the transmission of the $i$-hop node on the $j$-th chain over a long time interval [0, *Time*].

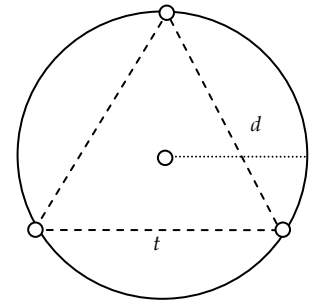

Fig. 7. Equilateral triangle inscribed in a circle.

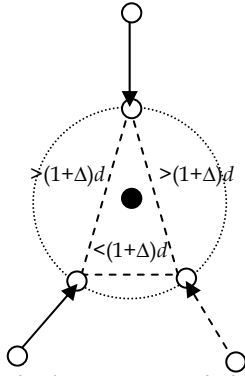

Fig. 8. At most two simultaneous transmissions from 2-hop nodes.

∎

Let $S_1$ = the union of airtimes occupied by all 1-hop nodes $S_{1j}$. Similarly, let $S_2$ = the union of airtimes occupied by all 2-hop nodes $S_{2j}$. That is, $S_1 = S_{11} \cup S_{12} \cup ... \cup S_{1N}$ and $S_2 = S_{21} \cup S_{22} \cup ... \cup S_{2N}$. We further define $x_{ij} = |S_{ij}|/Time$.

By definition,

$$|S_1 \cup S_2| \leq Time \qquad (5)$$

According to *assumption (3)*, when any 1-hop node transmits, none of the other 1-hop nodes or 2-hop nodes can transmit at the same time if collisions are not to happen. Thus, if carrier-sensing works perfectly and collisions due to simultaneous countdown to zero in the 802.11 backoff algorithm are negligible (see *assumptions (4)* and *(5)* in Section II), then

$$S_1 \cap S_2 = \varnothing \qquad (6)$$

and

$$S_{1i} \cap S_{1j} = \varnothing \qquad \text{for } i \neq j. \qquad (7)$$

This implies

$$|S_1| + |S_2| = |S_1 \cup S_2| \leq Time \qquad (8)$$

and

$$|S_1| = |S_{11}| + |S_{12}| + ... + |S_{1N}|. \qquad (9)$$

By *Corollary 1*,

$$|S_2| \geq \frac{|S_{21}| + |S_{22}| + ... + |S_{2N}|}{2}. \qquad (10)$$

Recall that we assume that the 1-hop nodes are relay nodes that do not generate data (see *Definition 5* and the justification before that in Section II). All traffic transmitted by 1-hop nodes must therefore come from 2-hop nodes. By the "no collision" assumption, the sum of the airtimes of 1-hop nodes must not be greater than the sum of airtimes of 2-hop nodes. We have

$$|S_{11}| + |S_{12}| + ... + |S_{1N}| \leq |S_{21}| + |S_{22}| + ... + |S_{2N}| \quad (11)$$

From (8)-(10), we have $|S_{11}| + |S_{12}| + ... + |S_{1N}| + (|S_{21}| + |S_{22}| + ... + |S_{2N}|)/2 \leq Time$. Applying (11), we get

$$(x_{11} + x_{12} + ... + x_{1N}) + \frac{(x_{11} + x_{12} + ... + x_{1N})}{2} \leq 1$$

giving

$$x_{11} + x_{12} + ... + x_{1N} \leq \frac{2}{3}$$

where $(x_{11} + x_{12} + ... + x_{1N})L$ is the throughput.

∎

We now show a specific schedule on a 2-chain network which achieves the capacity of $2L/3$. Consider the topology shown in Fig. 9. There are two chains, having link distance $d$ and *CSRange* = $2.9d$, which removes HN. Recall that the general HFD has two requirements, (i) RS mode and (ii) *CSRange* > $(3+\Delta)\ d_{max}$ [6]. For the topology in Fig. 9, it turns out that *CSRange* = $2.9d$ is enough.

The numbers shown on the links in Fig. 9 represent a possible transmission schedule. Links with same number transmits at the same time. Following this pattern, the throughput capacity of $2L/3$ is "potentially" achievable. Our simulation results in Subsection B below show that the 802.11 protocol throughput capacity is below but close to this upper bound.

The reader may be curious as to why we did not use a "symmetric" 2-chain network (where the angle between the chains is $\pi$) as the illustrating example above. It turns out that the symmetric structure cannot achieve the throughput of $2L/3$ if there are source nodes four or more hops away. To see this, first we note that for a symmetric 2-chain network, *CSRange* must be at least $3d$ to ensure HFD in the areas around the sink node (see discussion of the example in Fig. 3 in Section II). Given *CSRange*=$3d$, each of the chains (assuming a long chain with more than four hops (or five nodes)) cannot have throughput of $L/3$, as can be easily verified by analysis of one linear chain [7], [9].

Before going to the next subsection, we note that Theorem 1 actually applies not just to canonical networks (the proof does not require it), but general networks in which (i) all links are of the same length; and (ii) source nodes are two hops are more away from the center. In other words, the chains leading to the data center need not be straight-line linear chains. Thus, Theorem 1 can be stated more generally as Theorem 1' below:

*Theorem 1'*: For equal-link-length general networks, $C_{802.11} \leq 2L/3$, where $L$ is the link capacity.

*Proof*: Same as Theorem 1 since Lemma 1 and Corollary 1 apply to general networks with equal link length also.

### (2) Variable Link-Length Networks

In this subsection, we consider canonical networks in which the distance between adjacent rings can be varied (i.e., $d_0$, $d_1$,... may be distinct). With this assumption, the capacity is upper-bounded by $3L/4$. This is proved in

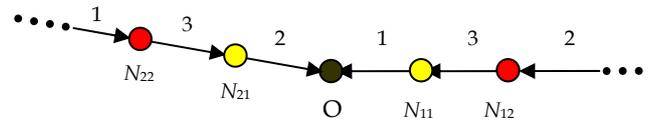

Fig. 9. Example of equal-link-length topology, *CSRange*=$2.9d$.

*Theorem 2* after *Lemma 2* in the following.

*Lemma 2*: At any time, at most *three* 2-hop nodes can transmit at the same time.

*Proof*: Assume the contrary that we can have four 2-hop nodes belonging to four different chains transmitting at the same time. With respect to Fig. 10, consider the four straight lines formed by the four nodes to the center (note: the network could have more chains, just that we are focusing on the four chains of the four 2-hop nodes in focus here). Four angles are formed between adjacent lines. Let $\theta < \pi/2$ be the minimum of the four angles. Four angles are also formed between non-adjacent lines. Let $\beta \leq \pi$ be the angle encompassing $\theta$ (see Fig. 10).

For simultaneous transmissions of 2-hop nodes, the transmitters should not be able to carrier-sense each other. This implies an upper bound for *CSRange* as follows:

$$CSRange < 2(d_0 + d_1)\sin\frac{\theta}{2} \ . \qquad (12)$$

In addition, by *assumption (5)*, to prevent collisions of 1-hop nodes and 2-hop nodes, they should be able to carrier-sense each other. This implies a lower bound for *CSRange*. By (4),

$$CSRange \geq \sqrt{(d_0+d_1)^2 + d_0^2 - 2d_0(d_0+d_1)\cos\beta} \ . \qquad (13)$$

By *assumption (3)*, the receivers of simultaneous transmissions should not violate the physical constraints. By (1),

$$(1+\Delta)d_1 < 2d_0 \sin\frac{\theta}{2} \ . \qquad (14)$$

Since there are four chains, $\theta \leq \pi/2$ and $\beta \leq \pi$. From the definitions of $\theta$ and $\beta$, we have

$$2\theta \leq \beta \leq \pi . \qquad (15)$$

From (13) and (15),

$$CSRange \geq \sqrt{(d_0+d_1)^2 + d_0^2 - 2d_0(d_0+d_1)\cos(2\theta)} \ . \qquad (16)$$

Let $d_1 = \alpha\, d_0$. We can form two inequalities from (12), (14) and (16):

$$\alpha < \frac{\sqrt{2(1-\cos\theta)}}{(1+\Delta)} \ , \qquad (17)$$

$$\begin{cases} \alpha > \dfrac{1-2\cos^2\theta}{1-2\cos\theta} + \dfrac{\sqrt{(2\cos^2\theta-1)^2 + 1 - 2\cos\theta}}{1-2\cos\theta} - 1 \\ or \\ \alpha < \dfrac{1-2\cos^2\theta}{1-2\cos\theta} - \dfrac{\sqrt{(2\cos^2\theta-1)^2 + 1 - 2\cos\theta}}{1-2\cos\theta} - 1 \end{cases} . (18)$$

Fig. 11 shows the plot of (17) and (18) when $\Delta = 0.78$. The shadowed region is the area of solution. From the plot,

$$\theta > 1.73 > \pi/2 .$$

This leads to a contradiction. Thus, there can be at most three simultaneous 2-hop transmissions.
∎

*Theorem 2*: For variable-link-length canonical networks, $C_{802.11} \leq 3L/4$, where $L$ is the link capacity.

*Proof*: Similar to the proof of *Theorem 1*, from *Lemma 2*,

$$|S_2| \geq \frac{|S_{21}| + |S_{22}| + ... + |S_{2N}|}{3} \ . \qquad (19)$$

Hence,

$$(x_{11} + x_{12} + ... + x_{1N}) + \frac{(x_{11} + x_{12} + ... + x_{1N})}{3} \leq 1$$

or $\quad x_{11} + x_{12} + ... + x_{1N} \leq \dfrac{3}{4}$ ,

where $(x_{11} + x_{12} + ... + x_{1N})L$ is the throughput
∎

Fig. 12 shows an example of a canonical network. The *CSRange* has to be set larger than $2.62d_0$ and smaller than $3.417d_0$. The numbers on the links show a possible transmission schedule that achieves capacity of $3L/4$.[1] Our simulation results in Subsection B below show that 802.11 throughput capacity is below but close to this upper bound.

In the analysis of canonical networks, we have assumed that the loss exponent is 4, corresponding to $\Delta =$

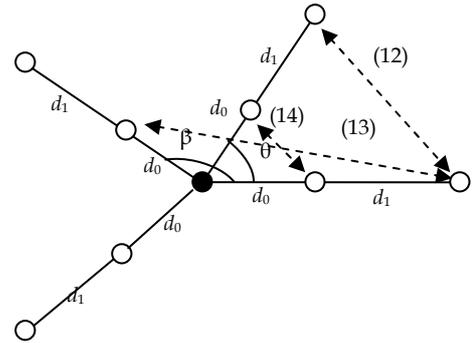

Fig. 10. Example of 4-chain canonical network.

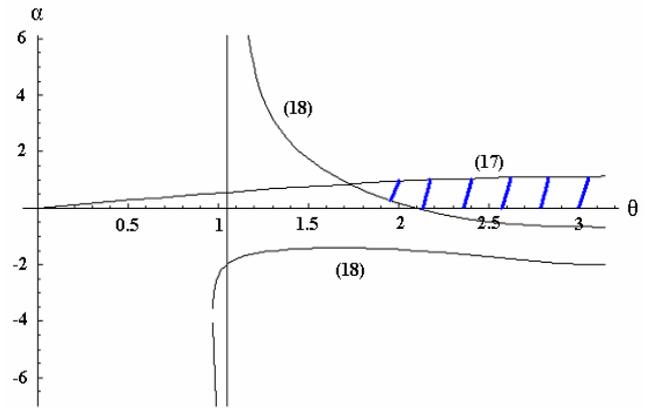

Fig. 11. Plot of Inequalities (17) and (18).

---

[1] For the one-to-many network (i.e., the sink becomes the source, and the sources become the sinks with respect to the many-to-one case here), some parameters should be changed to attain the capacity of $3L/4$. Specifically, *CSRange* = $1.7d_0$, and $d_i = 0.7d_0$ for i=1, 2, … The derivation method for the capacity of the one-to-many case is similar to that in the many-to-one case here.

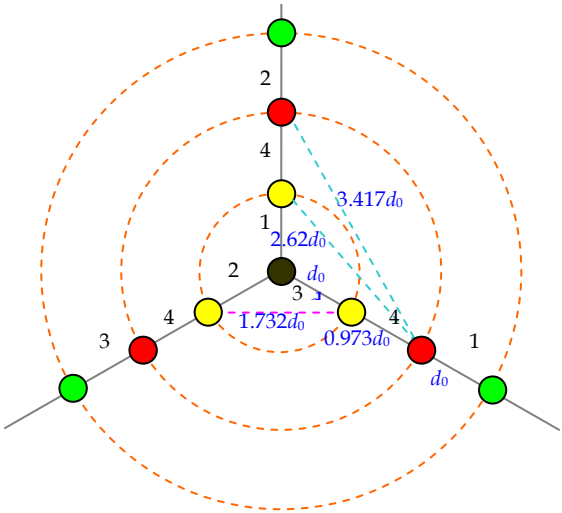

Fig. 12. Example of 3-chain canonical network, $CSRange$=2.7$d$.

0.78. In outdoor environment, the typical value of loss exponent is in the range 2 to 4. Similar analytical technique can be used to find their throughput capacities. Since smaller loss exponent implies larger $\Delta$ (larger interference), the throughput capacity under the assumption of loss exponent 4 serves as an upper-bound for the throughput capacity in outdoor environment.

### 3.2 Simulation

We use the network simulator NS2 [10] to simulate the canonical network shown in Fig. 12. As shown in Subsection A, for the 3-chain canonical network, $C_{802.11} \leq 3L/4$. In the simulation, the RS Mode is enabled. Table I shows the details of the simulated configuration. Only the $n$-hop nodes at the boundary are source nodes that generate data. Offered load control is applied to prevent them from injecting too much traffic into the network. For the interested reader, it has been shown in [7] that offered-load control can yield higher throughput in multi-hop networks.

Fig. 13 shows the simulation result assuming the set-up of Table I. The $x$-axis is the number of nodes per chain, including the sink. Given a number of nodes per chain, we vary the offered load in the simulation to identify an offered load that achieves the highest average throughput. When the number of nodes per chain is 3, i.e., the 2-hop nodes are the source nodes, the throughput is 4.62Mbps (0.740$L$), which is very close to the theoretical capacity 3$L$/4, where the link capacity $L$ is around 6.24Mbps as obtained by simulating one single link. But when the number of nodes per chain increases, the throughput drops to 4.30Mbps (0.690$L$).

An explanation for this phenomenon is that the scheduling scheme of IEEE 802.11 does not result in the optimal transmission schedule presented in Fig. 12 needed to achieve the 3$L$/4 upper bound. That is, the incorporation of random backoff countdown time in 802.11 causes imperfect scheduling. Consider Fig. 12, it is possible for 2-hop and 3-hop nodes of different chains to transmit at the same time in 802.11, since they are out of the carrier-sensing range of each other. To achieve capacity 3$L$/4, however, all the 2-hop nodes must transmit together. However, a 3-hop transmission may prevent this, resulting in only some of the 2-hop nodes transmitting together. In other words, there are times when not all 2-hop nodes transmit together, meaning $|S_2|$ cannot reach the lower bound in (19). Meeting the lower bound, however, is essential to achieving the optimal throughput 3$L$/4.

Fig. 14 shows the simulation results of canonical networks with different numbers of chains but with equal link length. The simulated configuration is shown in Tables II and III. For the 2-chain canonical network, we use the network structure in Fig. 9. The angle between two chains are slightly less than $\pi$. The reason of not using a symmetric structure has been given in Subsection A above. For other cases, the chains are evenly placed on the network. The CSRange for each topology is determined by minimizing its value while preventing HN. The throughput is obtained by varying the offered load and choosing the highest one. From the graph, the highest throughput is 3.86Mbps (0.619$L$), which is slightly smaller than the theoretical capacity of 2$L$/3. This is due to the imperfect scheduling by 802.11, which has been discussed in the previous paragraph.

In Fig. 14, the throughput converges to around 2.0Mbps (0.321$L$) when the number of chains increases. The convergence can be explained as follows. From the analysis in Subsection A, we see that the bottleneck is around the center. When the number of chains is large, the area near the center will become dense. The possible transmission patterns are similar in this area, and thus the throughput converges. In addition, note that the converged value, 0.321$L$, is considerably smaller than the value achieved when the number of chains is three, 0.619$L$. This is again due to imperfect scheduling of 802.11 MAC protocol. An interesting insight is that when the number of chains is small, the possible transmission patterns arise from "random" 802.11 MAC scheduling is more limited. And by limiting this degree of freedom, higher throughput can actually be achieved because random transmission patterns that degrade throughputs are eliminated.

The above observation has two implications: *(i) For network design, we may want to design the network in such a way that the number of routes leading to the center is limited. (ii) Even for a general, non-canonical, network densely populated with nodes and with many routes leading to the center, it is better to selectively turn on only a subset of the nodes to limit the routes to the center.* This principle will be further discussed in Section IV.

TABLE I
SIMULATION CONFIGURATION FOR VARIABLE-LINK-LENGTH CANONICAL NETWORKS

| Number of chains | 3 |
| --- | --- |
| $d_0$ | 250m |
| $d_1$ | 242m |
| $d_i$ for $i>1$ | 250m |
| Transmission Range | 250m |
| Carrier Sensing Range | 675m |
| Routing Protocol | AODV |
| Propagation Model | Two Ray Ground |
| Packet Data Size | 1460 bytes |

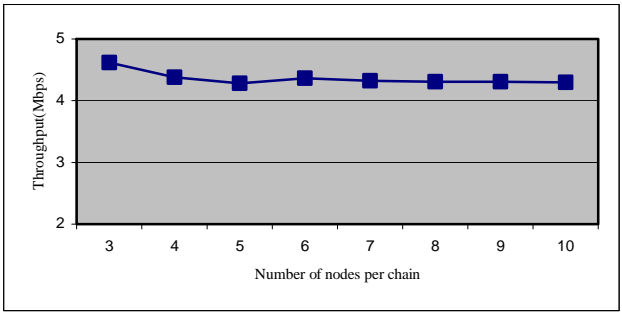

Fig. 13. Simulated throughput of a 3-chain canonical network with offered load control.

TABLE II
SIMULATION CONFIGURATION FOR EQUAL-LINK-LENGTH CANONICAL NETWORKS

| | |
|---|---|
| Number of nodes per chain | 8 |
| $d_i$ for all $i$ | 250m |
| Transmission Range | 250m |
| Carrier Sensing Range | Refer to Table III |
| Routing Protocol | AODV |
| Propagation Model | Two Ray Ground |
| Packet Data Size | 1460 bytes |

TABLE III
CARRIER SENSING RANGE FOR EQUAL-LINK-LENGTH CANONICAL NETOWKRS

| Number of chains | Carrier Sensing Range |
|---|---|
| 2 | 725m |
| 3 | 875m |
| 4 | 750m |
| 5 | 725m |
| 6 | 875m |
| 7 | 800m |
| 8 | 750m |
| 9 | 875m |
| 10 | 825m |
| >10 | 900m |

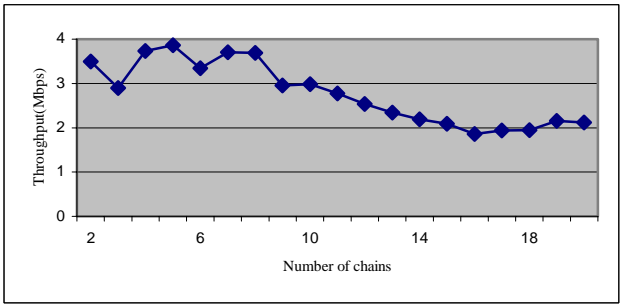

Fig. 14. Simulated throughput of equal-link-length canonical networks with offered load control.

### 3.3 HFD versus Non-HFD Performance

In the preceding sections, we have assumed HFD networks to simplify the analysis by eliminating the effect of collision. We now investigate the performance of HFD versus that of non-HFD networks. As a reminder, HFD requires

(i) Use of Receiver Restart (RS) Mode, and
(ii) Sufficiently large CSRange.

From [11], we know that increasing CSRange increases the number of exposed nodes (EN) and decrease the number of hidden nodes (HN), and vice versa. When HN is removed, say with HFD, the EN phenomenon will be more severe, which lowers the throughput. However, that is the case for many-to-many data delivery only. For this paper, we are interested in many-to-one data delivery. Table IV shows the simulation results with same configuration as in Table II with varying CSRange. The shaded entries correspond to HFD. From the table, when the number of chains is between 2 to 10, the highest throughput is achieved if we choose the smallest CSRange within HFD. This shows that the best HFD configuration generally works better than non-HFD.

TABLE IV
SIMULATION RESULT FOR EQUAL-LINK-LENGTH CANONICAL NETWOKRS

| Throughput (Mbps) | | No. of Chains | | | | | | | | |
|---|---|---|---|---|---|---|---|---|---|---|
| | | 2 | 3 | 4 | 5 | 6 | 7 | 8 | 9 | 10 |
| CSRange (m) | 975 | 2.388 | 2.981 | 3.355 | 2.833 | 2.863 | 3.022 | 2.891 | 3.054 | 3.114 |
| | 925 | 2.793 | 2.993 | 3.329 | 3.518 | 2.837 | 2.805 | 2.943 | 3.270 | 3.108 |
| | 875 | 2.797 | **2.999** | 3.508 | 3.535 | **3.393** | 3.272 | 3.163 | **3.384** | 2.883 |
| | 825 | 2.795 | 2.490 | 3.513 | 3.483 | 2.615 | **3.681** | 3.575 | 3.053 | **3.366** |
| | 775 | 2.808 | 2.473 | **3.724** | 3.540 | 2.760 | 2.754 | **3.709** | 3.367 | 3.269 |
| | 725 | **3.589** | 2.226 | 3.210 | **3.854** | 2.095 | 2.264 | 3.147 | 3.199 | 2.686 |
| | 675 | 3.170 | 2.288 | 2.398 | 2.799 | 2.142 | 2.261 | 2.176 | 2.367 | 2.633 |
| | 625 | 3.166 | 1.806 | 2.219 | 2.657 | 1.735 | 2.020 | 2.670 | 1.906 | 2.156 |
| | 575 | 3.183 | 1.788 | 2.168 | 2.202 | 1.657 | 1.609 | 2.280 | 1.929 | 2.041 |

bold: highest throughput; shaded: HFD

The better performance of HFD could be explained as follows. When CSRange is decreased, the number of HN increases and the number of EN decreases. More links could be active when there are fewer EN, thus the throughput in multiple-source-multiple-destination network could be higher in the non-HN free situation. In a many-to-one network, however, all the traffic is directed toward the same destination. With a non-HN free design, although the total throughput on a link basis (point-to-point throughput) may be increased, the many-to-one throughput (or the end-to-end throughput) could not benefit from the increase, because all the traffic in the end will flow toward the bottleneck and be dropped there due to HNs. We will see later that this observation suggests a design in which the area near the center should be made HN-free, while areas far away from the data center need not be HN-free.

### 3.4 Multiple Interference

Thus far, we have considered pair-wise interferences only. The analysis of pair-wise interferences is appealing from the simplicity viewpoint. However, it may not have taken into account the fact that the interferences from several other simultaneously transmitting sources may add up to yield unacceptable SIR even though each of the interferences may not be detrimental. In this section, we extend our analysis to take into account the effect of multiple interferences. For brevity, we refer to the throughput capacity obtained by assuming pair-wise interferences as pair-wise-interference throughput capacity, and the throughput capacity with mul-tiple interferences taken into account as multiple-interference throughput capacity.

The multiple-interference throughput capacity is in general less than or equal to that of the pair-wise throughput capacity. The question then is whether the pair-wise-inter-ference capacity is a tight bound for mul-tiple-interference capacity. We show in the following that this is indeed the case. In the following, we focus on the 3-chain network. The analytical argument and the qualitative results for the 2-chain network are similar.

Consider the canonical network in the Fig. 15, where $d_0=d_2=d_3=d_4$, and $d_1=0.9d_0$. In some cases, the SIR may not satisfy the constraint 10dB. For example, when $N_{11}$ is receiving DATA from $N_{12}$, and at the same time $N_{21}$ and $N_{31}$ are replying ACK to $N_{22}$ and $N_{32}$, the SIR is at most

$$\frac{P_{N_{11}}(N_{12})}{P_{N_{11}}(N_{21})+P_{N_{11}}(N_{31})} = \frac{\frac{P_t}{(0.9d)^4}}{\frac{P_t}{d^4}\left(\frac{1}{1.7321^4}+\frac{1}{1.7321^4}\right)} \approx 6.859$$

where $P_X(Y)$ is the received power from node Y to node X, $P_t$ is the transmission power.

This situation, however, occurs only if multiple ACKs are transmitted simultaneous in nearby links near the center. The probability of this occurring is low, since the transmission time of ACK is much lower than that of DATA. If we ignore the simultaneous transmissions of ACKs in these nearby links, we can show that the SIR due to multiple interferences is still more than 10dB, given that the SIR due to pair-wise interferences is more than 10dB, as follows.

*1. 1-hop node to sink node*

When the sink node is receiving DATA from $N_{11}$, the nearest three active links that cause largest interference are: $N_{23}$ to $N_{22}$, $N_{33}$ to $N_{32}$ and $N_{14}$ to $N_{13}$. If no two ACKs are transmitted simultaneously by these three links, the "worst-case" interference power at $N_0$ (which includes ACK from $N_{22}$ DATAs from $N_{33}$ and $N_{14}$, and transmissions by other nodes) is at most

$$P_{N_0}(N_{22})+P_{N_0}(N_{33})+P_{N_0}(N_{14})+P_{N_0}(N_{25})+P_{N_0}(N_{35})+P_{N_0}(N_{16})+...$$
$$=\frac{P_t}{d^4}\left(\frac{1}{1.9^4}+\frac{1}{2.9^4}+\frac{1}{3.9^4}+\frac{1}{4.9^4}+\frac{1}{4.9^4}+\frac{1}{5.9^4}+...\right) \approx 0.0995 \frac{P_t}{d^4}$$

Hence, the SIR is at least $1/0.09949=10.513$

*2. 2-hop node to 1-hop node*

Consider the link $N_{12}$ to $N_{11}$. The nearest three active links are: $N_{22}$ to $N_{21}$, $N_{32}$ to $N_{31}$, $N_{15}$ to $N_{14}$. Similar to above, the SIR is at least

$$\frac{P_{N_{11}}(N_{12})}{P_{N_{11}}(N_{21})+P_{N_{11}}(N_{32})+P_{N_{11}}(N_{15})+P_{N_{11}}(N_{24})+P_{N_{11}}(N_{34})+P_{N_{11}}(N_{17})+...}$$
$$=\frac{\frac{P_t}{(0.9d)^4}}{\frac{P_t}{d^4}\left(\frac{1}{1.7321^4}+\frac{1}{2.5515^4}+\frac{1}{3.9^4}+\frac{1}{4.4844^4}+\frac{1}{4.4844^4}+\frac{1}{5.9^4}+...\right)}$$
$$\approx 10.5259$$

*3. 3-hop node to 2-hop node and others*

The interference is less than the above cases. This part is skipped because the analytical approach is similar.

In the above, we have argued analytically the consid-eration of multiple interferences will not have substan-tially different performance than that of pair-wise inter-ference. We have focused on the 3-chain network with variable link distance because this structure provides the highest capacity bound among the canonical networks.

We now present simulation results for general canoni-cal networks with arbitrary number of chains. We have modified the NS2 simulator to take into account the ef-fects of multiple interferences (the modified NS2 code can be downloaded from the website in [12]). The throughput results are shown in Fig. 16. The multiple-interference throughput is only lower than the pair-wise-interference throughput by a small margin, and therefore the pair-wise-interference throughput serves a good bound for multiple-interference throughput.

## 4 GENERAL NETWORKS

In this section, we consider the throughput of general networks. Since general networks may not have the regu-lar structure of canonical networks, the throughput capac-ity could be lower than $3L/4$. We propose a method to find the capacity by selecting Hidden-node Free Paths (HFP).

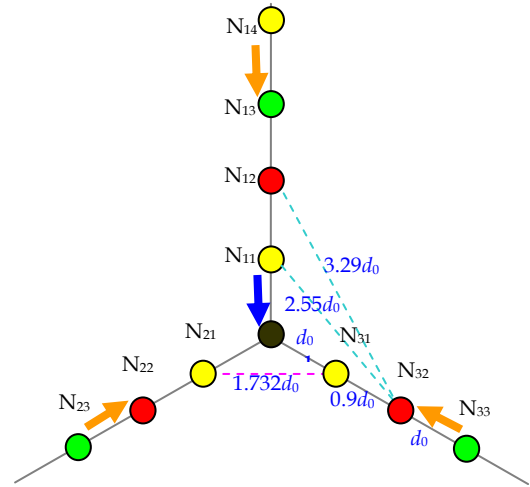

Fig. 15. Example of 3-chain canonical network, *CSRange=2.7d*.

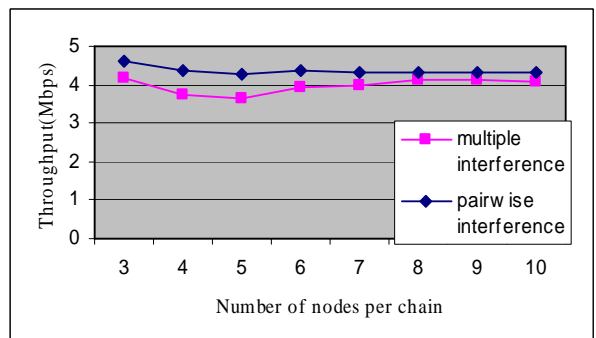

Fig. 16. Simulated throughput of 3-chain canonical network with offered load control.

## 4.1 Discussion of HFP

In Section III-C, we found that the network with HN-free outperforms that with HN in terms of throughput capacity. We could have three schemes which satisfy the HN-free condition for general network analysis. As one of the requirements of HFD, we assume RS Mode is used in all the analyses and experiments in the remaining of the paper. We assume that all nodes use a common fixed CSRange in each of the following schemes (assumption (1) in Section II); however, the schemes set the fixed CSRange differently.

*Scheme 1*: CSRange is set to $3.78 \cdot$ TxRange, where TxRange is the transmission range. This is a sufficient condition of HN free for any networks [6].

*Scheme 2*: CSRange is minimized according to the network topology so that no hidden node exists with respect to any two links in the network. This scheme, for example, was used in the analysis of canonical networks.

*Scheme 3*: HFP - We select a subset of links to form paths to the center which are hidden-node free and achieve the highest possible throughput. Since some links are not used, the CSRange can be smaller than scheme 1 and 2 (i.e., only the links in the path are considered when fixing CSRange.)

Based on Table IV, the highest throughput is achieved when we choose the smallest CSRange within HFD. So we have the following predictions for the throughputs of the different schemes above. The throughput of scheme 1 cannot be higher than that of scheme 2 (because the CSRange of some links are forced to adopt a higher value than necessary in scheme 1). Also, the throughput of scheme 2 cannot be higher than that of scheme 3 (because scheme 3 requires the HN property to be maintained only for links along the paths, and the paths that will be used are optimally chosen with regard to the throughput; whereas scheme 2 requires all links to be HN-free, even for links that are not used). For an example where HFP can achieve a higher throughput than scheme 2, we add two nodes to the 3-chain canonical network in Fig. 12 to yield the network in Fig. 17. In the network, link BB' interfere with link AA'. If we set CSRange to be less than $3.417d_0$, node B will become a hidden node of link AA'. If we set CSRange larger than $3.417d_0$, the capacity upper-bound $3L/4$ cannot be achieved. On the other hand, if we use HFP, we could select the links in the canonical network only. So node A could be "switched off" and there will not be hidden-node problem if we set CSRange to $2.7d_0$.

## 4.2 Experiments and Discussions

To conserve space, this paper will not go into the details of the formulation of the HFP problem, and the HFP experimental methodology. For the interested readers, such details can be found in the Appendix of our technical report [12]. In a nutshell, our approach extends that of [13] by additionally taking into consideration the effects of carrier sensing and HFD requirements. We also provide a branch-and-bound heuristic algorithm for the resulting integer linear program (ILP). Here we only present the performance results of experiments on schemes 1, 2, and 3 and their implications. Solving the ILP of scheme 3 is computationally intensive. The experimental results of scheme 3 in this subsection are therefore obtained using our branch-and-bound heuristic. Schemes 1 and 2 are still solved in an optimal manner. As will be seen, even with a suboptimal heuristic, scheme 3 still yields better results.

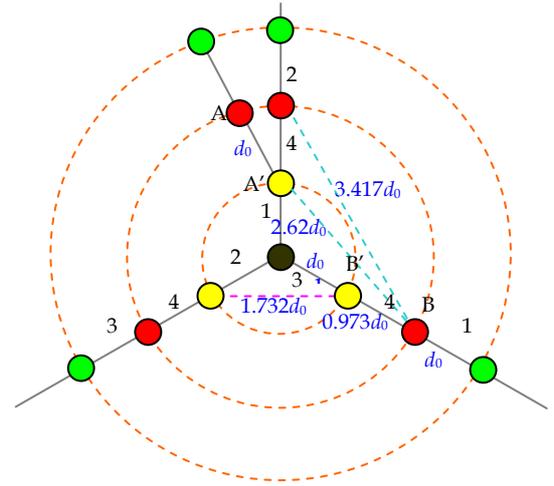

Fig. 17. Example of HFP.

In our experiments, we put the nodes inside a disk of radius one. A sink node is placed at the center of the disk, and six source nodes are placed evenly at the boundary of the disk spaced evenly apart. For each source node, a node is randomly generated within the transmission range 0.4. More nodes are generated similarly with reference to the newly created node until a node is within the transmission range from the sink node. In this way, we could ensure that there is a path from any source node to the sink node. By setting the transmission range to 0.4, the data from the source nodes will need at least three hops to reach the sink node.

Table V shows the experiment results for five randomly generated networks, Net$_1$, Net$_2$, …, Net$_5$. $T_1$, $T_2$ and $T_3$ are the throughputs of the three schemes. In obtaining $T_i$, we vary the offered load at the source nodes until the highest throughput is obtained [7]. From Table V, scheme 3 has improvements of 4.8% to 43.8% over scheme 1, and 4.8% to 23.2% over scheme 2. As related earlier, we did not solve scheme 3 optimally, but rather used a heuristic. Therefore, the CSRange (CS) found for HFP in the experiments may not be the shortest possible CSRange. Nevertheless, the result shows that the solutions of scheme 3 exhibit some properties similar to the canonical network, as shown in Fig. 12. We discuss the similarities in the following paragraph.

First, for scheme 3, CSRange/TxRange (CS$_3$/TX) for Net$_1$ to Net$_5$ is in the range of 2.62 to 3.417, which is the CSRange region we mentioned near the end of Section III-A for achieving the capacity of $3L/4$ in a canonical network. Second, exactly three paths leading to the sink node are used, which is the same as the 3-chain canonical network (Fig. 18). This gives us an intuition that the canonical network is in a sense optimal – that is, we may want to form a structure similar to the canonical network by turning on only some of the relay nodes.

TABLE V
RESULT FOR THROUGHPUT OF RANDOM NETWORKS

|  | $T_1$ | $T_2$ | $T_3$ | $T_3/T_1$ | $T_3/T_2$ | $CS_3$ | $CS_3/TX$ |
|---|---|---|---|---|---|---|---|
| Net$_1$ | 0.4 | 0.5 | 0.575 | 1.438 | 1.15 | 1.253 | 3.133 |
| Net$_2$ | 0.412 | 0.439 | 0.541 | 1.313 | 1.232 | 1.265 | 3.162 |
| Net$_3$ | 0.429 | 0.451 | 0.536 | 1.25 | 1.189 | 1.265 | 3.163 |
| Net$_4$ | 0.429 | 0.5 | 0.6 | 1.4 | 1.2 | 1.205 | 3.012 |
| Net$_5$ | 0.5 | 0.5 | 0.524 | 1.048 | 1.048 | 1.287 | 3.216 |

Net$_i$: Network $i$

TX: Transmission range, set to 0.4 in experiments

$T_1$: Throughput when CSRange=3.78 TX (Scheme 1)

$T_2$: Throughput when CSRange is minimized with respect to links in the network (Scheme 2).

$T_3$: Throughput when only some links in the network are activated (HFP) (Scheme 3)

$CS_3$: CSRange for Scheme 3

### 4.3 Applying Canonical Network to General Networks

The preceding subsection shows that HFP outperforms other HN-free schemes in terms of throughput. We also observe from the results that (i) HFP solutions for a random network exhibit structures similar to that of the 3-chain canonical network near the center. Furthermore, from simulation results in Section III-B (see Fig. 13), we observe that (ii) IEEE 802.11 scheduling in the canonical network achieves throughput close to that of perfect scheduling. Observations (i) and (ii) lead to the following general engineering principle:

Centric Canonical-Network Design Principle

- In a general multi-hop network densely populated with relay nodes, instead of solving the complex HFP optimization problem, as a heuristic, we may select routes near the center so that the structure looks like that of a 3-chain canonical network.

- If we have the freedom for node placement near the center during the network design process, then the nodes around the center should be structured like a 3-chain canonical network.

Note that there is no restriction on nodes far away from the center, and that they can be randomly distributed (see Fig. 19 for illustration).

This subsection investigates the application of the Centric Canonical-Network Design Principle. For our simulations, we assume there is a disk with radius 2000m. Within the disk, there is an inner circle with radius 980m. As illustrated in Fig. 19, the inner circle is structured as a canonical network. The nodes outside the inner circle are placed randomly with the constraint that the smallest distance between any two of them is not shorter than 125m. The nodes outside the inner circle act as source nodes and relay nodes at the same time, while the nodes inside the inner circle act merely as relay nodes. We refer to the network structure in Fig. 19 as *centric canonical network*, alluding to the fact that only the vicinity of the center looks like a canonical network. Henceforth, we shall refer to vicinity of the center as the canonical network and the randomly-structured part beyond that as the random network. The number of nodes beyond the inner circle is 284. We use the default setting in NS2, CSRange of 550m and TXRange of 250m, for performing the simulations. AODV routing is assumed. For the canonical network, with respect to Fig.12, we set $d_0$=200m. Since 550m/200m=2.75, which is within the range 2.62 to 3.417 (see Fig. 12), the canonical network is HN free. The random network, however, is not necessarily HN-free in our experiments. The assumption is reasonable, and corresponds to the real situation in which we only try to design the network architecture near the center judiciously by careful node placement.

As a benchmark, we have also conducted simulation experiments for a random network in which the inner circle is populated by 146 randomly placed nodes with no constraint on the node-to-node distance. In all our simulations below, the offered load to the source nodes are varied until we find the largest throughput for each network structure [7]. Simulation of 802.11 with AODV yields a throughput of 1.16 Mbps for the benchmark random network, and a throughput of 2.79Mbps for the centric canonical network. That is, the throughput of the centric canonical network is more than 100% higher. This demonstrates that *a carefully designed structured network around the data center yields superior performance*.

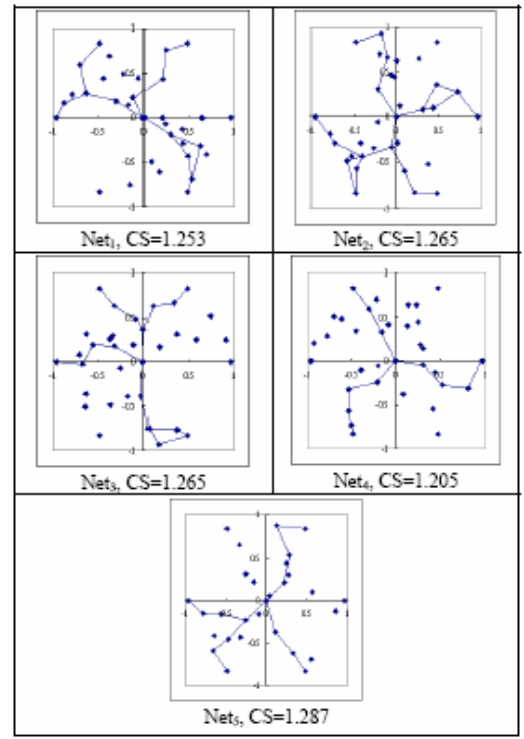

Fig. 18. Random Networks and HFP..

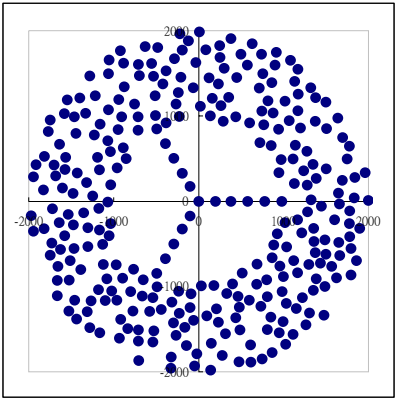

Fig. 19. Example of a centric-canonical.

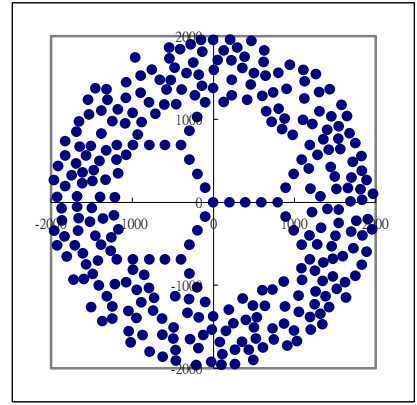

Fig. 20. Example of a manifold canonical network.

Although the improvement is significant, 2.79 Mbps is still a bit lower than the 4.30Mbps simulated throughput of the 3-chain canonical network in Section III. It turns out that the centric-canonical network actually fails to take another bottleneck into account. That is, in addition to the bottleneck around the center, there is also a bottleneck at the "confluence" of the random network and the canonical network, where the canonical network may branch off to many paths in the random network, and the nodes on these branches may interfere with each other in a negative way to bring down the throughput.

To mitigate the bottleneck at the confluence, we modify the canonical network as in Fig. 20. As shown, each chain in the canonical network only branches out further into two chains before meeting the random network. We refer to this design as the manifold canonical network, in reference to the fact that there are actually two "layers" of canonical networks. The first one is at the center, with three more before meeting the random network. We refer to this design principle as the *Manifold Canonical-Network Design Principle*.

In our simulations, the manifold canonical network is placed inside an inner circle of radius 1026s. The nodes beyond the manifold canonical network are randomly generated with the same constraints as the nodes generated beyond the inner circle of the centric canonical network. As the inner circle is larger than previous networks and the number of nodes (which are relay nodes) in the manifold canonical network is 31, to keep the total number of nodes in the network constant, the number of randomly generated nodes (which are also the source nodes) outside the inner circle is decreased from 284 to 269. We set CSRange 550m and $d_0$=200m in the manifold canonical network in our simulation (see Fig. 12). Simulation of 802.11 with AODV routing yields a throughput of 3.34Mbps, which is 20% higher than that of the centric canonical network. For fair bench-marking, we again perform the simulation with the inner circle replaced by random node placements, but this time with the inner circle having a radius of 1026m, as in the manifold canonical network. The simulation of the benchmark network yields a throughput of 1.31Mbps. We find that the throughput of the manifold canonical network is more than 150% over that by the pure random benchmark network.

We have also investigated the robustness of the manifold canonical network with respect node positioning. Simulations show that 5% position error of the nodes in the two "layers" of the canonical network only decreases the throughput by 10% on average, as summarized in TABLE VI.

TABLE VI
COMPARISON OF THROUGHPUTS OF MANIFOLD CANONICAL NETWORKS WITH AND WITHOUT NODE POSITION ERROR

| Throughput without position error (Mbps) | Throughput with position error (Mbps) | Ratio |
| --- | --- | --- |
| 3.44 | 3.45 | 1.003 |
| 3.35 | 3.11 | 0.928 |
| 3.32 | 3.18 | 0.958 |
| 3.29 | 2.94 | 0.894 |
| 3.37 | 2.96 | 0.878 |
| 3.36 | 2.82 | 0.839 |

## 5 CONCLUSION

In this paper, we have studied the throughput capacity of many-to-one multi-hop wireless networks based on the IEEE 802.11 MAC protocol. We have defined a class of *canonical networks* whose throughput capacity serves as a benchmark for general networks. Specifically, the throughput capacity of canonical networks under 802.11 is upper bounded by $3L/4$, where $L$ is the single-link capacity, when the source nodes are at least two hops away from the sink.

If we restrict our attention to networks in which all links have the same length, the upper bound is further reduced to $2L/3$. While the $3L/4$ result in the previous paragraph has been established for canonical networks only, the $2L/3$ result applies to general networks so long as (i) source nodes are at least two hops away from the data center; (ii) all links have the same length.

Our 802.11 simulation results yield throughputs are around $0.690L$ (for variable-link-length canonical networks) and $0.619L$ (for equal-link-length canonical networks) under the worse-case scenario when the source nodes are very far away and their traffic needs to go through many hops before reaching the sink node. That is, the simulated throughputs are reasonably close to the theoretical upper bounds of $3L/4$ and $2L/3$, respectively. This is a quite positive result considering the fact that 802.11 schedules transmissions in a rather random manner, while the examples we gave in Section III-A to achieve throughputs of $3L/4$ and $2L/3$ require very specific transmission orders.

The above results also imply that using variable link length is more desirable than using fixed link length. When the network is very dense (say, infinitely dense), if each node chooses a routing path with maximum hop distance in each hop, an equivalent network with fixed link length $d_{max}$, may result, where $d_{max}$ is the maximum hop-distance governed by the transmit power and receiver sensitivity. This max-hop-distance routing is not optimal for the many-to-one traffic pattern.

This paper has considered both canonical networks with and without hidden nodes. Our results indicate that hidden-node free designs (HFD) yield higher throughput capacity. This is in contrast to the many-to-many case where HFD may not yield better throughputs [5] [6] and may actually decrease the overall system throughput.

For general networks, we have used the concept of HFP (Hidden-node Free Path) to set up routes that yield optimal throughput. HFP routing, however, requires solving a complicated integer linear program, which may not be practical. Fortunately, our experimental results indicate that the routes selected by the HFP algorithm resemble the structure of the canonical network near the center. This gives rise to simple network design principles that attempt to approximate the canonical network structure in the center. Specifically, we have shown that a *manifold canonical network structure* near the sink can yield superior throughput that is as much as 150% higher than that of a dense random network.  A key insight is that in a network densely populated with nodes, deliberating turning off some nodes in the area near the sink node so as to approximate the canonical network structure can actually give rise to better throughput performance.

**Chi Pan Chan** received his B.Eng and M.Phil. degrees in Information Engineering from The Chinese University of Hong Kong in 2004 and 2006. His research was mainly related to capacity analysis in multi-hop wireless networks. He is now involved in the software industry in the field of multimedia and networking.

**Soung Chang Liew** received his S.B., S.M., E.E., and Ph.D. degrees from the Massachusetts Institute of Technology. From March 1988 to July 1993, Soung was at Bellcore (now Telcordia), New Jersey, where he engaged in Broadband Network Research. Soung is currently Professor and Chairman of the Department of Information Engineering, the Chinese University of Hong Kong. Soung's current research interests focus on wireless networking. Recently, Soung and his student won the best paper awards in the *1st IEEE International Conference on Mobile Ad-hoc and Sensor Systems (IEEE MASS 2004)* the *4th IEEE International Workshop on Wireless Local Network (IEEE WLN 2004)*. Separately, TCP Veno, a version of TCP to improve its performance over wireless networks proposed by Soung and his student, has been incorporated into a recent release of Linux OS. Publications of Soung can be found in www.ie.cuhk.edu.hk/soung. Besides academic activities, Soung is also active in the industry. He co-founded two technology start-ups in Internet Software and has been serving as consultant to many companies and industrial organizations. He is currently consultant for the Hong Kong Applied Science and Technology Research Institute (ASTRI), providing technical advice as well as helping to formulate R&D directions and strategies in the areas of Wireless Internetworking, Applications, and Services.

**An Chan** received the B.Eng degree in Information Engineering from The Chinese University of Hong Kong, Hong Kong in 2005. He is currently working toward a M.Phil degree in the same field at The Chinese University of Hong Kong. His research interests are in QoS over wireless network and advanced IEEE 802.11-like multi-access protocols. He is a graduate student member of IEEE.